# Fake News Detection via NLP is Vulnerable to Adversarial Attacks

Zhixuan Zhou[1,2], Huankang Guan[1], Meghana Moorthy Bhat[2] and Justin Hsu[2]

[1]*Hongyi Honor College, Wuhan University, Wuhan, China*
[2]*Department of Computer Science, University of Wisconsin-Madison, Madison, USA*
{kyriezoe, hkguan}@whu.edu.cn, {mbhat2, justhsu}@cs.wisc.edu

Abstract: News plays a significant role in shaping people's beliefs and opinions. Fake news has always been a problem, which wasn't exposed to the mass public until the past election cycle for the 45th President of the United States. While quite a few detection methods have been proposed to combat fake news since 2015, they focus mainly on linguistic aspects of an article without any fact checking. In this paper, we argue that these models have the potential to misclassify fact-tampering fake news as well as under-written real news. Through experiments on Fakebox, a state-of-the-art fake news detector, we show that fact tampering attacks can be effective. To address these weaknesses, we argue that fact checking should be adopted in conjunction with linguistic characteristics analysis, so as to truly separate fake news from real news. A crowdsourced knowledge graph is proposed as a straw man solution to collecting timely facts about news events.

## 1 INTRODUCTION

Fake news is an increasingly common feature of today's political landscape. To help address this issue, researchers and media experts have proposed fake news detectors adopting natural language processing (NLP) to analyze word patterns and statistical correlations of news articles. While these detectors achieve impressive accuracy on existing examples of manipulated news, the analysis is typically quite shallow—roughly, models check whether news articles conform to standard norms and styles used by professional journalists. This leads to two drawbacks.

First, these models can detect fake news only when they are under-written, for instance when the content is totally unrelated to the headline (so-called "clickbait") or when the article includes words considered to be biased or inflammatory. While this criteria suffices to detect many existing examples of fake news, more sophisticated rumor disseminators can craft more subtle attacks, for instance taking a well-written real news article and tampering the article in a targeted way. By preserving the original subject matter and relating the content tightly to the headline without using biased phrases, an *adversarial* article can easily evade detection. To demonstrate this kind of attack, we evaluate a state-of-the-art model called Fakebox. We introduce three classes of attacks: fact distortion, subject-object exchange and cause confounding. We generate adversarial versions of real news from a dataset by McIntire (2018), and show that Fakebox achieves low accuracy when classifying these examples.

At the same time, requirements posed by current detectors are often too strict. Real news which is under-written or talks about certain political and religious topics is likely to be mistakenly rejected, regardless of its accuracy. This is a particularly serious problem for open platforms, such as Twitter in the United States and TouTiao in China, where much of the news is contributed by users with diverse backgrounds. To prevent frustrating false positives, platforms are still heavily relying on manual work for separating fake news from real news. We provide experimental evidence for Fakebox's potential of misclassifying real news.

Taken together, our experiments highlight vulnerable aspects of fake news detection methods based purely on NLP. Without deeper semantic knowledge, such detectors are easily fooled by fact-tampering attacks and can suffer from a high rate of false positives, mistakenly classifying under-written yet real news which may not be written in a journalistic style. To address these problems, we argue that some form of fact-based knowledge must be adopted alongside NLP-based models. What this knowledge is remains to be seen, but we consider a straw man solution: a crowdsourced knowledge graph that aggregates infor-mation about news events and helps judge whether information extracted from news articles is reliable.

Table 1: Examples of fact tampering attacks.

| Attack type | Original | Adversarial |
|---|---|---|
| Fact distortion | 12 people were injured in the shooting. | 24 people were killed in the shooting. |
| Subject-object exchange | A gangster was shot by the police. | A policeman was shot by the gangster. |
| Cause confounding | The condom policy originated in 1992 ... The Boy Scouts have decided to accept people who identify as gay and lesbian. (unrelated events) | The inclusion of gays, lesbians and girls in the Boy Scouts led to the condom policy. |

The rest of our paper is organized as follows. In Section 2, we introduce three kinds of attacks targeting fake news detectors. Section 3 introduces the dataset from which we generate adversarial examples and our target model Fakebox. We evaluate Fakebox's performance under attacks in Section 4. In Section 5, we discuss weaknesses of NLP-based detectors and propose augmenting detectors with knowledge information, such as a crowdsourced knowledge graph, as a step towards making fake news detection more robust. Finally, we discuss related work in Section 6 and conclude in Section 7.

## 2 ADVERSARIAL ATTACKS

*Adversarial Machine Learning* is an emerging field of applied machine learning that seeks to understand how machine learning classifiers can be attacked by malicious users. To see how well existing fake news detectors perform against adversarial inputs, we explored three kinds of adversarial examples with tampering focusing on different aspects of an article:

- Fact distortion: exaggerating or modifying on some words. Character, time, location, relation, extent and any other element can be distorted;

- Subject-object exchange: with this attack readers will be confused as to who is the performer and who is the receiver of an action. It can be performed on sentence level;

- Cause confounding: either building non-existent causal relationship between two independent events, or cutting off some parts of a story, leaving only the parts that an adversarial wants to present to his readers.

Examples of these attacks are shown in Table 1. By repeating these modifications, we can significantly change the semantic content of a news article without distorting the "writing style" of the original one—the modified article is still presented in a seemingly logical and sound way.

## 3 DATASET AND MODEL

### 3.1 Dataset

We generate adversarial examples from articles in McIntire's fake-real-news-dataset, an open-source dataset extensively used in misinformation research. The dataset contains 6,335 articles. 3,171 of them are labeled as real and 3,164 of them are labeled as fake. The ratio of real and fake news articles is roughly 1:1. Titles, contents and veracity labels are provided. The dataset does not include URLs, but we are primarily concerned with the textual content rather than external links, which can be manipulated in many other ways. We manually check veracity of the news by comparing them with reputable sources to increase our confidence that the labels are reasonable.

### 3.2 Fakebox

Fakebox analyzes linguistic characteristics of news articles to assess whether they are likely to be real news or not. By looking at different aspects of an article (title, content and URL), using NLP models and training on a manually curated database, Fakebox can successfully identify fake news. Edell (2018) reports achieving classification accuracy upwards of 95%.

Fakebox checks several aspects of each article:

- Title or headline: checked for clickbait;

- Content: analyzed to determine whether it's written like real news;

- Domain: some websites are known for hosting certain types of content, like hoaxes and satires.

If an article is written like a real one, Fakebox labels it as impartial and gives it a score between 60 and 100. If an article is not written like a real one, Fakebox labels it as biased and gives it a score between 0 and

Table 2: Normal-time output of Fakebox.

| Labels | Impartial | Biased | Unsure |
|---|---|---|---|
| Real news | 1159 | 1477 | 535 |
| Fake news | 537 | 2184 | 443 |

Table 3: Normal-time accuracy of Fakebox with unsure cases excluded.

| News type | Number of articles | Correctly classified | Classification accuracy |
|---|---|---|---|
| Real | 2636 | 1159 | 43.97% |
| Fake | 2721 | 2184 | 80.26% |
| Total | 5357 | 3343 | 62.40% |

40. Otherwise, Fakebox labels it as unsure and gives it a score between 40 and 60. It labels and assigns quantitative scores for titles, contents and domains, respectively. Higher-scoring articles are likely to be more reliable.

## 4 EXPERIMENTAL EVALUATION

The main focus of Fakebox is on linguistic characteristics of a news article without any fact checking, which potentially makes it vulnerable when facing news which is written in a similar style to real news but is not factual. To test this hypothesis, we performed an experimental evaluation of Fakebox. We establish a baseline by testing Fakebox with unmodified examples from McIntire's dataset. Then, we apply our attacks described in Section 2. All experiments are conducted on an Intel machine equipped with quad-core 1.80 GHz CPU, 8GB RAM, 256GB SSD and running Windows 10.

### 4.1 Baseline Performance

We first test baseline performance of Fakebox with McIntire's dataset. We feed 6,335 headlines and articles into Fakebox and get back corresponding labels. We take special care of its output labels for content veracity. Output of Fakebox is shown in Table 2. Real and Fake are actual attributes of news articles while Impartial, Biased and Unsure are labels given by Fakebox. We use true positive (TP) to denote correctly classified fake news, true negative (TN) to denote correctly classified real news, false positive (FP) to denote misclassified real news and false negative (FN) to denote misclassified fake news. False positive rate (FPR) and false negative rate (FNR) are defined as below:

$$FPR = FP/(FP+TN)$$
$$FNR = FN/(FN+TP)$$

Fakebox's accuracy on McIntire's dataset is 52.77%, false rate is 31.79% and for the other 15.44% samples, Fakebox is unsure about their veracity. While it is acceptable for fake news detectors to be unsure for some articles and leave the hard tasks to field experts, which is what happens in news platforms moderation nowadays, its accuracy in our experiment is still unsatisfactory even if we don't take unsure labels into consideration. It performs well when dealing with fake news where false negative rate is only 19.74%. But on the other hand, it labels more real news as biased than as impartial. Quantitatively, its false positive rate is 56.03%. Its overall accuracy when excluding unsure cases is 62.40%. The result is shown in Table 3.

We observe that in false positive cases, words that tend to be regarded as "fake" include "anti", "prison", "terror", "Islamism" and "Trump". This focus on sensitive terms leads to a crude analysis. Though much fake news emerges around these topics, it is not appropriate to give a large "bias weight" to these words, which is implemented in many state-of-the-art models. After all, there are equally many real news articles talking about these issues.

We also observe that many of the real articles misclasified as fake can be regarded as under-written, i.e., they are not written in a journalistic style. They are likely to be written by grass-root writers from Twitter who are interested in social issues and are willing to share their opinions instead of professional journalists.

## 4.2 Attack-time Performance

We generate adversarial examples by hand from real news that are also labeled as impartial by Fakebox. This process can be seen as an adversary selecting real news from authority sources like New York Times and performing considerable tampering by himself.

For fact distortion, we simply substitute people, places or actions without much effort. For example, for the article titled "Is the GOP losing Walmart?", we substitute each "Walmart" in the content with "Apple". The veracity score given by Fakebox drops down by only 0.0073, which is negligible for its judgement. Tampering to other articles as well doesn't cause the veracity score to drop by much. This kind of tiny tampering can have an outsized impact—imagine that company A is involved in an information breach scandal but company B is reported to be responsible by fake news.

For subject-object exchange tampering, the veracity score doesn't change at all, since term frequency stays the same. This can be quite misleading: "a gangster was shot by the police" and "a policeman was shot by the gangster" are totally different and the latter will cause public panic.

Cause checking is probably the most vulnerable part of NLP-based detectors. For instance, there are two real and unrelated articles labeled as impartial by Fakebox, one about Walmart scoring 0.7151 for veracity and the other about local politics in Cleveland scoring 0.7652. When we simply mix the two articles together, the generated article is still labeled as impartial and even reaches a much higher score (0.8585). We further try to mix an article labeled as impartial with an article labeled as biased and the veracity score for the generated article is between the scores of two original articles, which indicates that only linguistic characteristics are inspected and facts are never checked. As long as an adversary keeps articles in a classical manner, he can mix totally unrelated events together, build non-existent causal relationships and evade detection.

## 5 DISCUSSION

As we can see in experiments, simply looking into linguistic aspects is not enough for fake news detection. Two main defects of this method are its vulnerability to fact tampering attacks and its bias towards under-written articles and certain topics. Given that one of the essences of fake news is fact tampering, fact checking could be quite helpful. However, it is largely missing in current fake news detection models.

## 5.1 A Straw Man Solution

There is urgent need to compare information extracted from news articles with "fact", and source of the fact is another key issue. Media and specialists have delicate skills yet limited time and energy to collect various fact from all sources. Fake news usually comes on early stage after events happen and thus requires early detecting, worsening the situation. One possible solution to stopping fake news is to extract key information from articles including causal relationships and compare it with a dynamically-updated news fact knowledge graph. Such an approach was also proposed by Pan et al. (2018).

A knowledge graph is a graph with entities of different types as nodes and various relations among them as edges (Jia et al., 2016). Typical examples include WordNet (Miller, 1995) and OpenKN (Jia et al., 2014) and realistic applications include document understanding (Wu et al., 2012) and link prediction (Liu et al., 2014). Knowledge Graph is also used by Google to enhance its search engine's results with information gathered from a variety of sources. The information is presented to users in an infobox next to the search results. An example Google knowledge graph is shown in Figure 1.

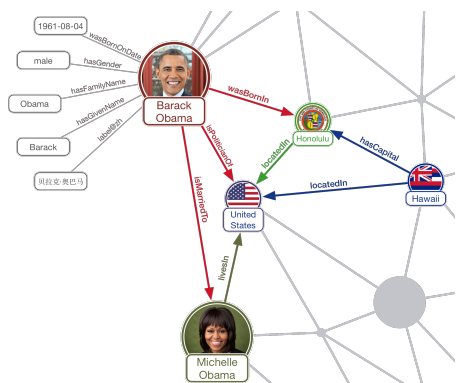

Figure 1: Example knowledge graph.

Crowdsourcing is a distributed problem-solving model in which a crowd of undefined size is engaged to solve a complex problem through open calls (Chatzimilioudis et al., 2012). It divides work between participants to achieve a cumulative result. It is possible that a large crowd of non-experts can collaborate well on a task that otherwise would require extensive efforts of a small group of experts (Howe, 2006). The news aggregation site Reddit.com (Mieghem, 2011) is another example of this method's application other than Wikipedia. While crowdsourcing tends to result in high disagreement among contributors, Dumitrache et al. (2018) showed that disagreement is not noise but signal, and that in fact crowdsourcing can not only be cheaper and scalable, it can be higher quality and more informative as well, for disagreement representation can be used to detect low quality workers.

A crowdsourced generation of knowledge graphs may be efficient and timely in the context of news propagation. While fake news usually floods on the early stage after an event happens, local or well-informed people hear about the events faster and more accurately. They can either be journalists or bystanders who are equally responsible for fact maintaining and fake news combating. If we can create a structured visualized interface for building and editing knowledge graphs, where users only need to fill in the "subject", "action", "object", "time" and "location" entities, they can easily fill in facts they are sure of without much professional expertise. The design could be visually similar to the Google knowledge graph shown in Figure 1—which is friendly to non-expert users—but it could work in a crowdsourced manner. As the knowledge graph is updated dynamically, timely fact information can be utilized to detect fact tampering attacks in news articles.

The main drawback of our straw man solution is the difficulty of collecting high-quality information. While the crowdsourcing way of updating the knowledge graph does ensure high efficiency, attackers with special intentions have equal access to creating and editing. If the fact entries collected with the knowledge graph are not fact but "accomplices" created by attackers, they cannot be utilized to help detect fake news. How to address this issue is a serious challenge when it comes to crowdsourcing.

## 5.2 What is a Fact?

It is highly difficult to give a clear and flexible definition of "fact requirement" for different information propagation contexts. In this work, we see "fact" from a conventional, qualitative perspective and understand it as a statement that generally conforms to a certain event or a piece of knowledge. Wikipedia defines a fact as something that is consistent with objective reality or that can be proven with evidence—if a statement can be demonstrated to correspond to experience, it's a fact. However, we recognize that this is by no means the end of the story and significantly more research is needed to make the idea of a fact more concrete.

## 6 RELATED WORK

Fake news detection became a hot research field in 2015. Since then, a number of methods have been put forward. We list early, foundational works including categorization of tasks and methods as well as building of platforms and datasets. We also list fake new detection methods in three main categories.

## 6.1 Foundational Works

Rubin et al. (2015) separated the task of fake news detection by type of fake: serious fabrications, large-scale hoaxes and humorous fakes.

Conroy et al. (2015) provided a typology of veracity assessment methods emerging from two major categories—linguistic cue approaches and network analysis approaches. They saw promise in an innovative hybrid approach that combined the two methods.

Shao et al. (2016)) introduced Hoaxy, a platform for the collection, detection, and analysis of online misinformation, which had the potential to help people understand the dynamics of real and fake news sharing. Wang (2017) presented LIAR, a publicly available dataset for fake news detection.

## 6.2 Linguistic Approaches

These models look simply at linguistic characteristics such as grammar feature, word pattern, term count and appearance of certain expressions. Defects of the models are discussed in detail in our paper.

Chen et al. (2015) examined potential methods for the automatic detection of clickbait. Methods for recognizing both textual clickbaiting cues and non-textual ones including image and user behavior were surveyed.

Bourgonje et al. (2017) presented a system for detecting the stance of headlines with regard to their corresponding article bodies. The approach could be applied in fake news, especially clickbait detection scenarios.

Granik and Mesyura (2017) showed a simple approach for fake news detection using naive Bayes classifier and achieved decent result considering the relative simplicity of their model.

Horne and Adali (2017) analyzed difference of fake and real news in title features, complexity and style of content. Elaboration Likelihood Model was considered as a theory to explain the spread and persuasion of fake news.

Rashkin et al. (2017) compared language of real news with that of satire, hoaxes, and propaganda to find linguistic characteristics of untrustworthy text.

Their experiments adopted stylistic cues to help determine the truthfulness of text.

## 6.3 Network Approaches

Models based on network analysis realize the importance of taking various background information into account, instead of inspecting solely the articles themselves. They perform generally well at most times, but when related information is missing or little, their performance will drop.

Jin et al. (2016) improved news verification by mining conflicting viewpoints in microblogs.

Long et al. (2017) proved that speaker profiles such as party affiliation, speaker title, location and credit history provided valuable information to validate the credibility of news articles.

Farajtabar et al. (2017) proposed a multi-stage intervention framework that tackled fake news in social networks by combining reinforcement learning with a point process network activity model.

Volkova et al. (2017) found social interaction features were more informative for finer-grained separation between four types of suspicious news (satire, hoaxes, clickbait and propaganda) compared to syntax and grammar features.

Tacchini et al. (2017) classified Facebook posts as hoaxes or non-hoaxes with high accuracy on the basis of the users who liked them.

## 6.4 Hybrid Approaches

Hybrid approaches combine the advantage of linguistic models and network models, which intuitively outperform either of them.

Ruchansky et al. (2017) proposed a model that combined the text of an article, the user response it receives, and the source users promoting it for a more accurate and automated prediction.

As far as we know, no other hybrid approaches are available and fact-checking is absent from all existing models. We also survey on commercial fake news detectors and find that the majority of them take only linguistic features into consideration.

## 7 CONCLUSION

In this paper, we evaluate a fake news detector Fakebox on adversarial attacks, including fact-distortion, subject-object exchange and cause confounding attacks. Experiments show that our attack subverts the model significantly. We believe that similar models based solely on linguistic characteristics will perform much less effectively in the real world and are especially vulnerable to tampering attacks. This kind of attack is much more subtle, since it doesn't change the overall writing style of news articles and thus has the potential to evade similarity detection. We argue that multi-source fact comparing and checking must be integrated into fake news detection models to truly detect misinformation.

At the same time we find false positive rate rises when it comes to either under-written real articles or certain topics around which there is supposed to be more fake news. The potential of misclassifying under-written yet real news will hurt amateur news writers' enthusiasm. Thus we further suggest using fact-checking as a helpful supplement so as to smooth the negative effect of false positive judges.

One possible way to collect fact about news events is to use a crowdsourced knowledge graph, which is dynamically updated by local and well-informed people. The timely information collected can then be used to compare to that extracted from news articles and help generate a label of veracity.

Our future work includes building a visualized interface for news knowledge graph crowdsourcing, so as to make work as easy as possible for non-experts and stop fact-tampering fake news on early stage. We also want to look at the issue of fake news propagation from a different angle, i.e., putting it in a social context and examining human factors in order to better understand the problem.